\documentclass[prd,twocolumn,aps,noshowpacs,nofootinbib,amsmath,amssymb,floatfix,superscriptaddress]{revtex4}
\usepackage[colorlinks=true,linkcolor=red,citecolor=blue]{hyperref}
\usepackage{amsmath}
\allowdisplaybreaks[4]
\usepackage{amsfonts}
\usepackage{graphicx}
\usepackage{subfigure}
\usepackage{dcolumn}
\usepackage{bm}
\usepackage{booktabs}
\usepackage[utf8]{inputenc}
\usepackage{multirow}
\usepackage{graphicx,graphics,dcolumn,booktabs,bm}
\usepackage{longtable,lscape}
\usepackage{txfonts}
\usepackage{overpic}
\usepackage{amssymb}
\usepackage{indentfirst}
\usepackage{epsfig}
\begin{document}

%\title{ Beauty-charmed meson $B_{c}$ decay constants matching at two-loop accuracy  }
\title{ Next-to-next-to-leading order matching of beauty-charmed meson $B_{c}$ and $B^*_{c}$ decay constants}
\author{ Wei Tao\footnote{taowei@njnu.edu.cn}}
%\author{ Mingqi Cao\footnote{@njnu.edu.cn}}
\author{ Ruilin Zhu~\footnote{Corresponding author: rlzhu@njnu.edu.cn} }
\author{Zhen-Jun Xiao \footnote{Corresponding author: xiaozhenjun@njnu.edu.cn} }

\affiliation{ Department of Physics and Institute of Theoretical Physics, Nanjing Normal University, Nanjing, Jiangsu 210023, China}
\date{\today}
\vspace{0.5in}

%\date{\today}
\begin{abstract}
We present the next-to-next-to-leading order (NNLO) QCD corrections to the decay constants for both the
pseudoscalar and vector beauty-charmed mesons $B_{c}$ and $B^*_{c}$ in nonrelativistic QCD effective theory.
Explicit NNLO calculation verified that the $B_c$ decay constant from pseudoscalar current is identical with
the $B_c$ decay constant from axial-vector current. The NNLO result for the vector decay constant of $B^*_{c}$ meson
is novel. Combined with the latest extraction of nonrelativistic QCD long-distance matrix elements of $B_c$ meson,
we give the branching ratios of leptonic decays of
$B_{c}$ and $B^*_{c}$ mesons. In addition, the novel anomalous dimension for the flavor-changing heavy quark vector current
in nonrelativistic QCD effective theory are helpful to investigate the threshold behaviours of two different heavy quarks.

 \vspace{0.3in}

%\begin{description}
%\item[PACS numbers] 12.38.Bx,  13.25.Gv, 14.40.Pq
%\item[Keywords]
% 12.38.Bx Perturbative calculations
% 13.25.Gv decays of jpsi,Upsilon and other quarkonia
% 14.40.Pq  heavy quarkonia
%\end{description}
\end{abstract}

\maketitle

\section{Introduction}

The beauty-charmed meson $B_{c}(1S)$ was first discovered in proton anti-proton colliders by CDF collaboration in 1998~\cite{CDF:1998ihx}.
The second new member in beauty-charmed meson family, i.e. $B_c(2S)$,  was discovered in proton proton colliders by ATLAS collaboration in 2014~\cite{ATLAS:2014lga}.
Five years later, the  $B_c(2S)$ state was confirmed by both CMS and LHCb collaborations, in addition, the new vector member $B^*_c(2S)$
 was first reported by  these two collaborations~\cite{CMS:2019uhm,LHCb:2019bem}. Up to now, no other member in beauty-charmed meson family is observed in particle physics experiment though
 more beauty-charmed mesons have been predicted in many theoretical models.

Unlike the heavy quarkonium, the experimental measurements of beauty-charmed meson family are not easy since they are composed of two different heavy flavor quarks and
the ground state $B_{c}(1S)$ only weak decays into other lighter particles. Though there are 48 decay channels listed in latest review of particle physics which have
been reported in experiments, no one has an experimental measurement of absolute branching ratios\footnote{An exception is the absolute branching ratio of $B_c^+\to \chi_{c0}\pi^+$, which is extracted by particle data group after inputting the bottom quark fragmentation probability into B meson and the LHCb data. } ~\cite{Workman:2022ynf}.

To promote the determination of the absolute branching ratios, it is required to carefully investigate the fundamental properties of decay behaviours.
In other words, we need first to have a good knowledge of the decay constants for the beauty-charmed meson family. In principle, the decay constants for the beauty-charmed mesons
are nonperturbative yet universal physical quantities. In this point Lattice QCD shall be a good method to determine the decay constants from the first principal of QCD.
However, the Lattice QCD studies on the beauty-charmed mesons are lesser because the beauty-charmed mesons include two different heavy quarks and the doubly heavy quark systems are not easy to be simulated in current lattices\footnote{There  is a $2\sigma$ tension for the $B_c$ decay constant between ETM lattice result and HPQCD lattice result~\cite{Becirevic:2018qlo,Colquhoun:2015oha}. Based on heavy highly improved staggered quark approach, HPQCD has also performed lattice QCD simulations on  the vector and axial-vector form factors of $B_c\to J/\psi$~\cite{Harrison:2020gvo}. }.

The nonrelativistic QCD (NRQCD) effective theory provides a systematical and accurate framework to study the doubly heavy quark systems~\cite{Bodwin:1994jh}. In this effective theory,
the heavy quark mass provides a nature factorization scale. The short-distance physics above the heavy quark mass can be perturbatively calculated and factorized into the Wilson coefficients while the long-distance physics below the heavy quark mass go into the long-distance matrix elements (LDMEs).
Within NRQCD effective theory, the decay constants for beauty-charmed mesons can be further factorized as the short-distance matching coefficients and the corresponding
NRQCD LDMEs.

Using NRQCD effective theory, the next-to-leading order (NLO) corrections including both the strong coupling constant correction at order $\alpha_s$ and relative velocity correction at order $v^2$
to the axial-vector decay constant of the $B_c$ meson  was first calculated by Braaten and Fleming in 1995~\cite{Braaten:1995ej},
after a systematical study of the $B_c$ meson at the leading order (LO) by Chang and Chen~\cite{Chang:1992pt}.
Using the resummation technique,  the NLO corrections including all order relative velocity corrections
to the axial-vector decay constant of $B_c$ meson  and the vector decay constant of $B^*_c$ was estimated by Lee, Sang, and Kim in 2010~\cite{Lee:2010ts}.
The next-to-next-to-leading order (NNLO) corrections at order $\alpha^2_s$ to the axial-vector decay constant of $B_c$ meson was first investigated by
Onishchenko and Veretin in 2003~\cite{Onishchenko:2003ui}. However, the full analytical expression of the axial-vector decay constant of $B_c$ meson at NNLO accuracy
was accomplished by Chen and Qiao in 2015~\cite{Chen:2015csa}. Very recently, the numerical calculation of the axial-vector decay constant of $B_c$ meson at
 NNNLO accuracy was by Feng, Jia, Mo, Pan, Sang, and Zhang~\cite{Feng:2022ruy}. Other higher-order calculation on doubly heavy quark system and phenomenological studies on $B_c$ system
 can be found, for example,  in the  literatures~\cite{Marquard:2006qi,Egner:2021lxd,Beneke:1997jm,Marquard:2009bj,Marquard:2014pea,Kniehl:2002yv,Egner:2022jot,Sang:2022kub,Feng:2022vvk,Chen:2022vzo,
 Chen:2017soz,Tao:2022yur,Tang:2022nqm,Zhu:2017lwi,Zhu:2017lqu,Zhao:2022auq,Bordone:2022drp,Sun:2022hyk,Xiao:2013lia,Wang:2012kw,Piclum:2007an}.

In this paper, we will calculate the  pseudoscalar decay constant of  $B_c$ meson  and the vector decay constant of $B^*_c$ meson at NNLO accuracy within NRQCD
effective theory. By an explicit calculation, we can investigate the relation among various decay constants defined by different flavor-changing heavy quark currents.
The pseudoscalar decay constant of  $B_c$ meson  is identical to the axial-vector decay constant of $B_c$ meson.
The NNLO results of the vector decay constant of $B^*_c$ meson  are novel. Combined with the latest extraction of the NRQCD LDMEs,  we give the branching ratios of
 leptonic decays of
$B_{c}$ and $B^*_{c}$ mesons.
 These results of matching coefficients are also useful to analyze the threshold behaviours when two different heavy quark
are close to each other.

In addition, we obtain a novel anomalous dimension for the flavor-changing heavy quark vector current at NNLO accuracy of NRQCD effective theory.
This anomalous dimension is related to the renormalization behaviours of the vector current with two different heavy quarks in NRQCD.

The paper is arranged as follows. In Sec. II, we give the definition of the decay constants from
pseudoscalar, axial-vector, and vector currents for beauty-charmed mesons $B_{c}$ and $B^*_{c}$ in both the full QCD theory and the NRQCD effective theory. We then present the matching formulae for the decay constants in  the NRQCD effective theory.  In Sec. III, we present
the calculation methods and calculation procedures for the short-distance matching coefficients. In Sec. IV, we give the final NNLO results of the short-distance matching coefficients
and the decay constants of $B_{c}$ and $B^*_{c}$.   We also perform a phenomenological analysis of the leptonic decays of $B_{c}$ and $B^*_{c}$.
We conclude in the end of the paper.

\begin{figure*}[th]
\begin{center}
\includegraphics[width=0.6\linewidth]{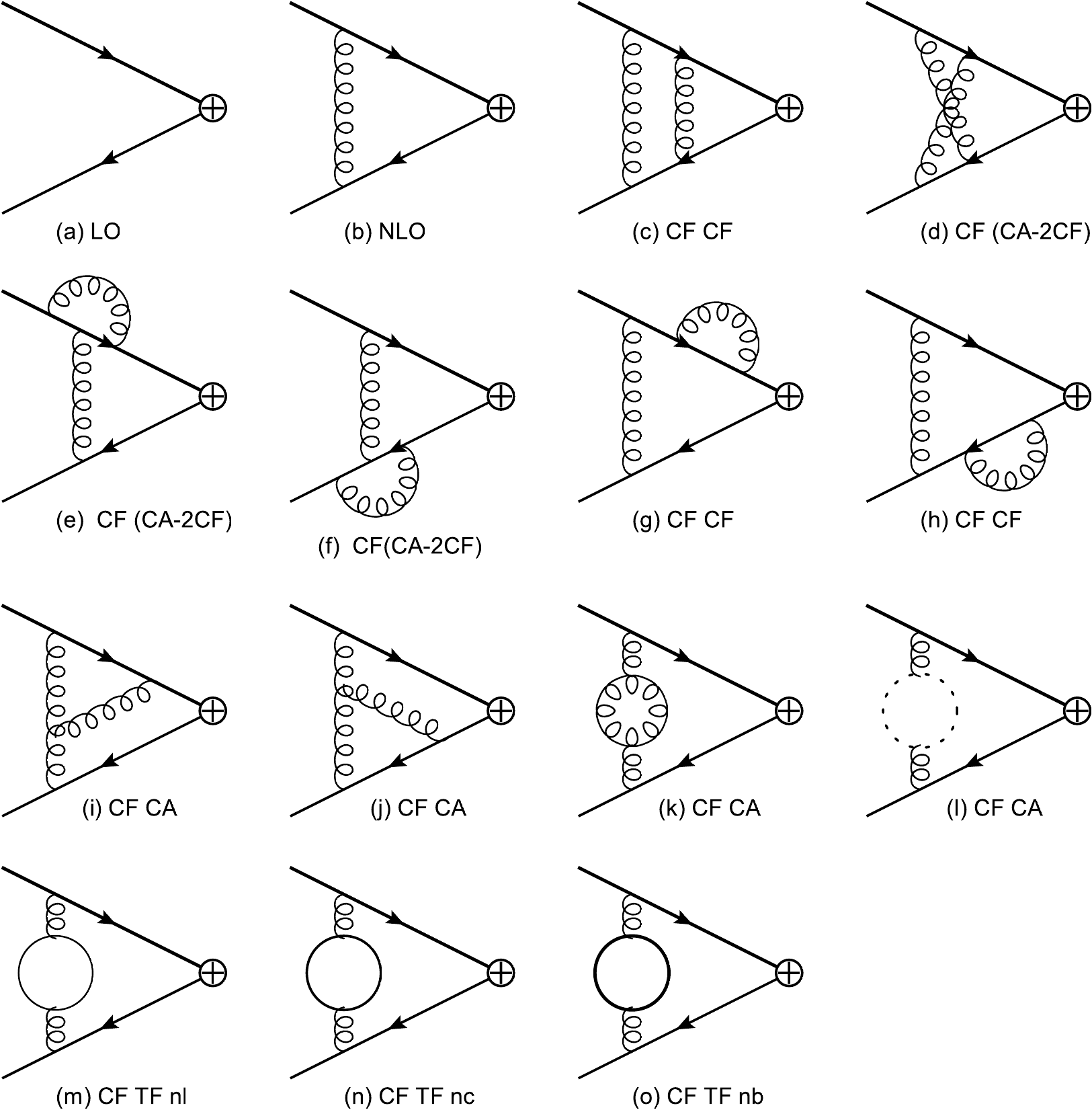}
\caption{The Feynman diagrams labelled with corresponding color factor  for $B_c$ and $B^*_c$ decay constants up to  two-loop order. The cross ``$\bigoplus$'' implies the insertion of certain heavy flavor-changing  current. The thinnest, thick, thickest solid circles represent $n_l$ massless quark-loop, $n_c$ quark-loop with mass $m_c$, $n_b$ quark-loop with mass $m_b$, respectively. In this paper, $n_b=n_c=1$.}\label{fig:bcwtree1loop2loop}
\end{center}
\end{figure*}

\section{Matching formulae }

Though the $B_c$ meson leptonic decay is dominated by the virtual $W$ boson with a $V-A$ weak interaction in particle physics standard model, one can freely define
 the $B_c$ meson decay constants by different flavor-changing currents. Thus one can define the pseudoscalar and vector $B_c$ meson decay constants by the full QCD
 matrix elements
 \begin{align}
\left\langle 0\left|\bar{b} \gamma^\mu \gamma_5 c\right| B_c(P)\right\rangle &=i f^a_{B_c} P^\mu,\\
\left\langle 0\left|\bar{b} \gamma_5 c\right| B_c(P)\right\rangle &=i f^p_{B_c}m_{B_c},\\
\left\langle 0\left|\bar{b} \gamma^\mu c\right| B^*_c(P,\varepsilon)\right\rangle &= f^v_{B^*_c} m_{B^*_c}\varepsilon^\mu,
\end{align}
where $\left|B_c(P)\right\rangle$ and $\left|B^*_c(P,\varepsilon)\right\rangle$  are  respectively the states of pseudoscalar and vector $B_c$ mesons with four-momentum $P$ and $\varepsilon^\mu$ is the polarization vector for  vector $B_c$ meson. In full QCD, the standard covariant normalization of the hadron state is $\left\langle B_c\left(P^{\prime}\right) \mid B_c(P)\right\rangle=(2 \pi)^3 2 P^0 \delta^3\left(P^{\prime}-P\right)$. The imaginary unit in the right hand of equations is added to make sure the $f_{B_c}$ being real and positive. Note that other decay constants for $B_c$ family with scalar and tensor currents are not considered in this paper. Using
the heavy quark equation of motion, one can easily get the identity  $f^a_{B_c}=f^p_{B_c}$. Thus we only need to calculate two decay constants  $f^p_{B_c}$ and $f^v_{B^*_c}$ in the following.

The above decay constants of $B_c$ mesons are principally nonperturbative observables in full QCD and rely on a nonperturbative calculation, however, the two heavy quark system is not well-simulated at current Lattice QCD and these physical quantities are rarely investigated in the first principal theory of QCD.

In NRQCD effective theory, the decay constants of $B_c$ mesons can be further factorized into a perturbatively calculable short-distance coefficients with the
corresponding nonperturbative LDMEs. Thus one can write the following matching formula at leading-order in relative velocity expansion
\begin{align}\label{formula}
 f^p_{B_c}&=\sqrt{\frac{2}{m_{B_c}}}\mathcal{C}_p(m_b,m_c,\mu_f)\left\langle 0\left|\chi^\dag_b \psi_c \right| B_c(\mathbf{P})\right\rangle(\mu_f)+{\cal O}(v^2),\\
 f^v_{B^*_c}&=\sqrt{\frac{2}{m_{B^*_c}}}\mathcal{C}_v(m_b,m_c,\mu_f)\left\langle 0\left|\chi^\dag_b \mathbf{\sigma}\psi_c \right| B^*_c(\mathbf{P})\right\rangle(\mu_f)+{\cal O}(v^2),
\end{align}
where $\mu_f$ is the NRQCD factorization scale which appears in the short-distance coefficients at two-loop calculation and will be cancelled between the short-distance coefficients and NRQCD LDMEs. In QCD perturbative calculation, the decay constants  will depend on the renormalization scale in fixed-order accuracy  and will become
renormalization scale independence after summing all-order contributions.

\section{Calculation of the matching coefficients }

In this section, we present our calculation procedures for the decay constants of pseudoscalar and vector $B_c$ mesons within NRQCD approach.
According to the above matching formulae, the matching coefficients $\mathcal{C}_p$ and $\mathcal{C}_v$ can be obtained by the calculation of both full QCD matrix elements
and the NRQCD matrix elements. At leading-order, the  matching coefficients $\mathcal{C}_p$ and $\mathcal{C}_v$ are set as
$\mathcal{C}_p=\mathcal{C}_v=1$, which can also be done after  the nonrelativistic expansion of heavy quark current.
The Feynman diagrams  for $B_c$ and $B^*_c$ decay constants up to  two-loop order are plotted in Fig.~\ref{fig:bcwtree1loop2loop}.

Our higher order calculation of the matching coefficients consists of the following steps.
First, we use {\texttt{FeynCalc}}~\cite{Shtabovenko:2020gxv}  to obtain Feynman diagrams and corresponding Feynman amplitudes. By {\texttt{\$Apart}}~\cite{Feng:2012iq}, we decompose  every Feyman amplitude  to several Feynman integral families. Second, we use {\texttt{Kira}}~\cite{Klappert:2020nbg}/{\texttt{FIRE}}~\cite{Smirnov:2019qkx}/{\texttt{ FiniteFlow}}~\cite{Peraro:2019svx} based on Integration by Parts(IBP)~\cite{Chetyrkin:1981qh} to reduce every Feynman integral family to master integral family. Third,
based on symmetry among different integral families and using  {\texttt{Kira}}+{\texttt{FIRE}}+{\texttt{Mathematica\,code}}, we can realize  integral reduction among different integral families, and further on, the reduction from all of master integral families to the minimal master  integral families. Last, we use {\texttt{AMFlow}}~\cite{Liu:2022chg}, which is a proof-of-concept implementation of the auxiliary mass flow method~\cite{Liu:2017jxz}, equipped with {\texttt{Kira}}/{\texttt{FiniteFlow}} to calclate the minimal master integral families one by one.

In order to obtain the high-order coefficient $\mathcal{C}_J$ with $J=p,~v$, one has to perform the  conventional renormalization procedure, which are similar to  what are shown in Refs.~\cite{Chen:2015csa,Kniehl:2006qw,Bonciani:2008wf,Davydychev:1997vh}, i.e., $Z_J Z_{2,b}^{\frac{1}{2}} Z_{2,c}^{\frac{1}{2}} \Gamma_J=\mathcal{C}_J\tilde{Z}^{-1}_J \tilde{Z}_{2,b}^{\frac{1}{2}} \tilde{Z}_{2,c}^{\frac{1}{2}} \tilde{\Gamma}_J$ where the left part in the equation represents the renormalization of full QCD current while the right part represents the renormalization of NRQCD current.
$Z_J$  and $\tilde{Z}^{-1}_J$  are the renormalization constants for full QCD and NRQCD flavor-changing currents, respectively. Here, $Z_a=Z_v=1$, $Z_p=\frac{m_b Z_{m,b}+m_c Z_{m,c}}{m_b+m_c}$. And $\tilde{Z}_{2,b}=\tilde{Z}_{2,c}=\tilde{\Gamma}_J=1$. Equivalently, We can also use diagrammatic renormalization method~\cite{deOliveira:2022eeq}, which contain two loop diagrams and three  kinds of counter-term diagrams, i.e.,  tree diagram   inserted with one $\alpha_s^2$-order counter-term vertex,  tree diagram   inserted with two $\alpha_s$-order counter-term vertexes (vanishing), and one loop diagram inserted with one $\alpha_s$-order counter-term vertex.

We want to mention that all contributions have been evaluated for general gauge parameter $\xi$ and the final results for the matching coefficients are all independent of $\xi$, which constitutes an important check on our calculation. In the calculation of two loop diagrams, we allow for one $b$ quark, one $c$ quark and $n_l$ massless quarks in the quark loop. Up  to two-loop order, the most complicated renormalization constants are the on-shell mass and wave function renormalization constants  allowing for two different non-zero quark masses~\cite{Bekavac:2007tk, Fael:2020bgs}, which are presented in the appendix.

After  renormalization,  the results of the short-distance matching coefficients $\mathcal{C}_J$ can
be expressed as
\begin{align}\label{C0express}
& \mathcal{C}_J(\mu_f,\mu,m_b,x)=
\nonumber\\&
1+\frac{\alpha_s^{(n_f)}}{\pi} \mathcal{C}_J^{(1)}(x) +\left(\frac{\alpha_s^{(n_f)}}{\pi}\right)^2
\left(\mathcal{C}_J^{(1)}(x)\frac{\beta_0^{(n_f)}}{4}\text{ln}\frac{\mu^2}{m_b^2}
+\frac{\gamma_{J}^{(2)}(x)}{2}\ln \frac{\mu_f^2}{m_b^2}
\right.\nonumber\\&\left.
+C^2_F \mathcal{C}^{FF}_J(x)+C_F C_A  \mathcal{C}^{FA}_J(x)
%\right.\nonumber\\&\left.
+C_F  T_F n_l \mathcal{C}^{FL}_J(x)+C_F  T_F \mathcal{C}^{FH}_J(x) \right)
\nonumber\\&
+\mathcal{O}(\alpha^3_s).
\end{align}
where we have defined a dimensionless parameter $x$ representing the ratio of two heavy quark masses
\begin{align}
&x= {m_c\over m_b}.
\end{align}
Besides we have suppress the renormalization scale $\mu$ dependence in strong coupling constant $\alpha_s$. The first two coefficients in $\beta$ functions for $\alpha_s$ are
\begin{align}
&\beta_0^{(n_f)}=(11/3)C_A-(4/3) T_F n_f,
\\&
\beta_1^{(n_f)}=(34/3)C_A^2-(20/3) C_A T_F n_f-4 C_F T_F n_f.\end{align}

If keeping the heavy quark mass in gluon self energies Feynman diagrams, the heavy quark mass will go into the running of the strong coupling constant. We apply the following decoupling relation~\cite{Chetyrkin:2005ia,Bernreuther:1981sg,Barnreuther:2013qvf,Grozin:2007fh,Ozcelik:2021zqt} to translate $\alpha_s^{(n_f)}$ involving massive flavours to  $\alpha_s^{(n_l)}$ only involving  $n_l$ massless flavours,
\begin{align}
\frac{\alpha_s^{(n_f)}}{\alpha_s^{(n_l)}}=1+\frac{\alpha_s^{(n_l)}}{\pi} T_F \left(\frac{n_b}{3} \text{ln}\frac{\mu^2}{m_b^2}+\frac{n_c}{3} \text{ln}\frac{\mu^2}{m_c^2}+\mathcal{O}(\epsilon)\right)+\mathcal{O}(\alpha^2_s),
\end{align}
where $n_f=n_l+n_b+n_c$.
In our numerical calculation,  $n_b=n_c=1$, $n_l=3$ and the two loop result
for strong coupling constant~\cite{Chetyrkin:1997sg,Chetyrkin:2000yt,Deur:2016tte,Herren:2017osy} is used, i.e.,
\begin{align}\label{aslexpress}
\alpha_s^{(n_l)}\left(\mu\right)=\frac{4\pi}{\beta_0^{(n_l)} \text{ln}\frac{\mu^2}{{\Lambda_{QCD}^{(n_l)}}^2}}
\left(1-\frac{\beta_1^{(n_l)}\text{ln}\,\text{ln}\frac{\mu^2}{{\Lambda_{QCD}^{(n_l)}}^2}}
{{\beta_0^{(n_l)}}^2 \text{ln}\frac{\mu^2}{{\Lambda_{QCD}^{(n_l)}}^2}}\right),
\end{align}
where  the typical QCD scale $\Lambda_{QCD}^{(n_l=3)}=336\mathrm{MeV}$  can be iteratively determined by $\alpha_s^{(n_l=5)}(m_Z)=0.1179$ with $m_Z=91.1876\mathrm{GeV}$.

Explicit analytical calculation of the NLO Feynman diagrams give  the NLO short-distance matching coefficients
\begin{align}
& \mathcal{C}_p^{(1)}(x)=\frac{3}{4}C_F\left(\frac{x-1}{x+1}\text{ln} x-2\right),
\\&
\mathcal{C}_v^{(1)}(x)=\frac{3}{4}C_F\left(\frac{x-1}{x+1}\text{ln} x-\frac{8}{3}\right).
\end{align}
Note that the analytical expressions of $\mathcal{C}_p^{(1)}(x)$  and $\mathcal{C}_v^{(1)}(x)$ are consistent with previous literatures~\cite{Braaten:1995ej,Lee:2010ts}.

At NNLO, the direct results of the matching coefficients are still IR-divergent after performing the UV renormalization.
 This is due to the UV divergence in the NRQCD LDMEs at NNLO.
 The anomalous dimensions $\gamma_{J}$ is related to $\tilde{Z}_J$ by
\begin{align}
\gamma_{J}=\frac{\text{d}\ln \tilde{Z}_{J}}{\text{d}\ln\mu}|_{\epsilon=0}=
\left(\frac{\alpha_s^{\left(n_f\right)}}{\pi}\right)^2 \gamma_{J}^{(2)}(x)+\mathcal{O}(\alpha^3_s),
\end{align}
whose solution gives the   renormalization constants $\tilde{Z}_J$ for different NRQCD currents
\begin{align}
\tilde{Z}_{J}=1-\left(\frac{\alpha_s^{\left(n_f\right)}}{\pi}\right)^2
\left( \frac{\mu^2}{\mu_f^2}\right)^{2\epsilon} \frac{\gamma_{J}^{(2)}(x)}{4\epsilon}+\mathcal{O}(\alpha^3_s),
\end{align}
where
\begin{align}
&\gamma_{p}^{(2)}(x)=-\pi^2\left(\frac{C_F C_A}{2}+\frac{(1+6x+x^2)C_F^2}{2(1+x)^2}\right),
\\&
\gamma_{v}^{(2)}(x)=-\pi^2\left(\frac{C_F C_A}{2}+\frac{(3+2x+3x^2)C_F^2}{6(1+x)^2}\right).
\end{align}
Note that the anomalous dimension $\gamma_{v}^{(2)}(x)$ is a novel result for two different heavy quarks meson. In the case of  $x=1$, the result is consistent with the previous calculation, for example in Ref.~\cite{Kniehl:2006qw}.
By the renormalization of the the UV divergence in the NRQCD LDMEs
 at NNLO, the extra IR-divergences in short-distance coefficients can be exactly cancelled. Thus we finally get the finite results for the matching coefficients.
\begin{figure}[thb]
	\centering
	\includegraphics[width=0.45\textwidth]{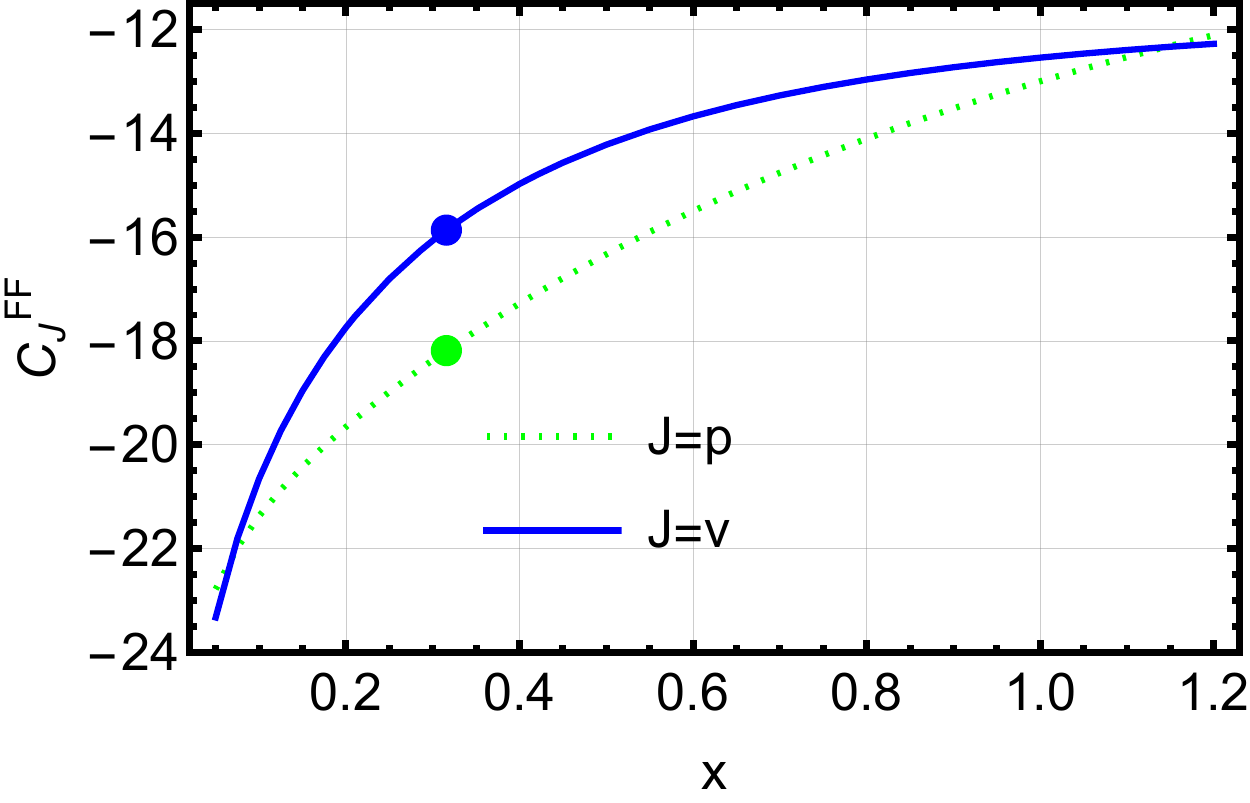}
	\caption{The profile of sub-coefficient $\mathcal{C}^{FF}_{J}(x)$ dependence on heavy quark mass ratio $x=\frac{m_c}{m_b}$ with $x\in [0.05, 1.2]$. $J=p$ represents the sub-coefficient for the pseudoscalar current while
$J=v$ represents the sub-coefficient for the vector current.
 According to  color/flavor structure, this sub-coefficient has a pre-factor $C_F^2$  for both the pseudoscalar current  and the vector current. The green and blue dots correspond to the results at physical heavy quark mass ratio with $x_0=\frac{1.5}{4.75}$. }
	\label{fig:CFFx}
\end{figure}

\begin{figure}[thb]
	\centering
	\includegraphics[width=0.45\textwidth]{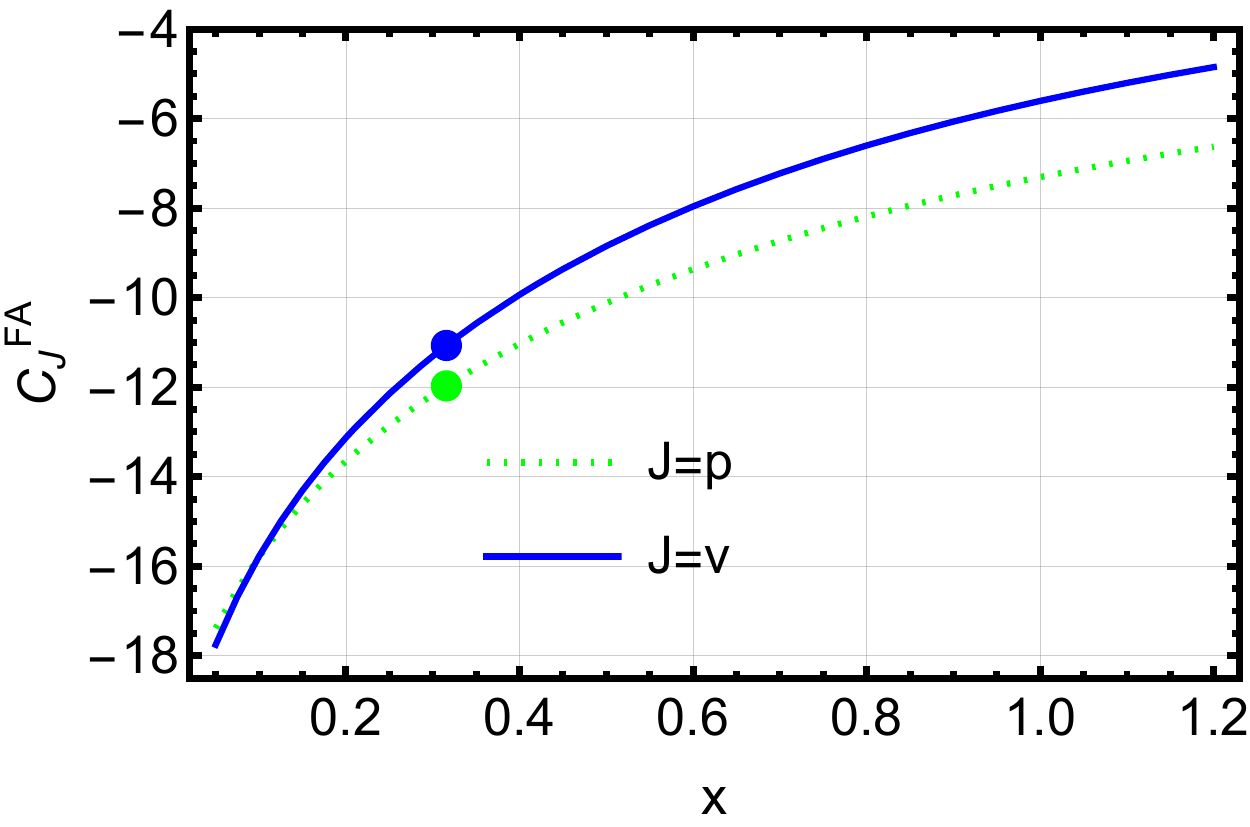}
	\caption{The same as Fig.~\ref{fig:CFFx}, but for  the profile of sub-coefficient $\mathcal{C}^{FA}_{J}(x)$ which has a pre-factor $C_F C_A$. }
	\label{fig:CFAx}
\end{figure}

\section{Numerical results and discussions}

\begin{figure}[thb]
	\centering
	\includegraphics[width=0.45\textwidth]{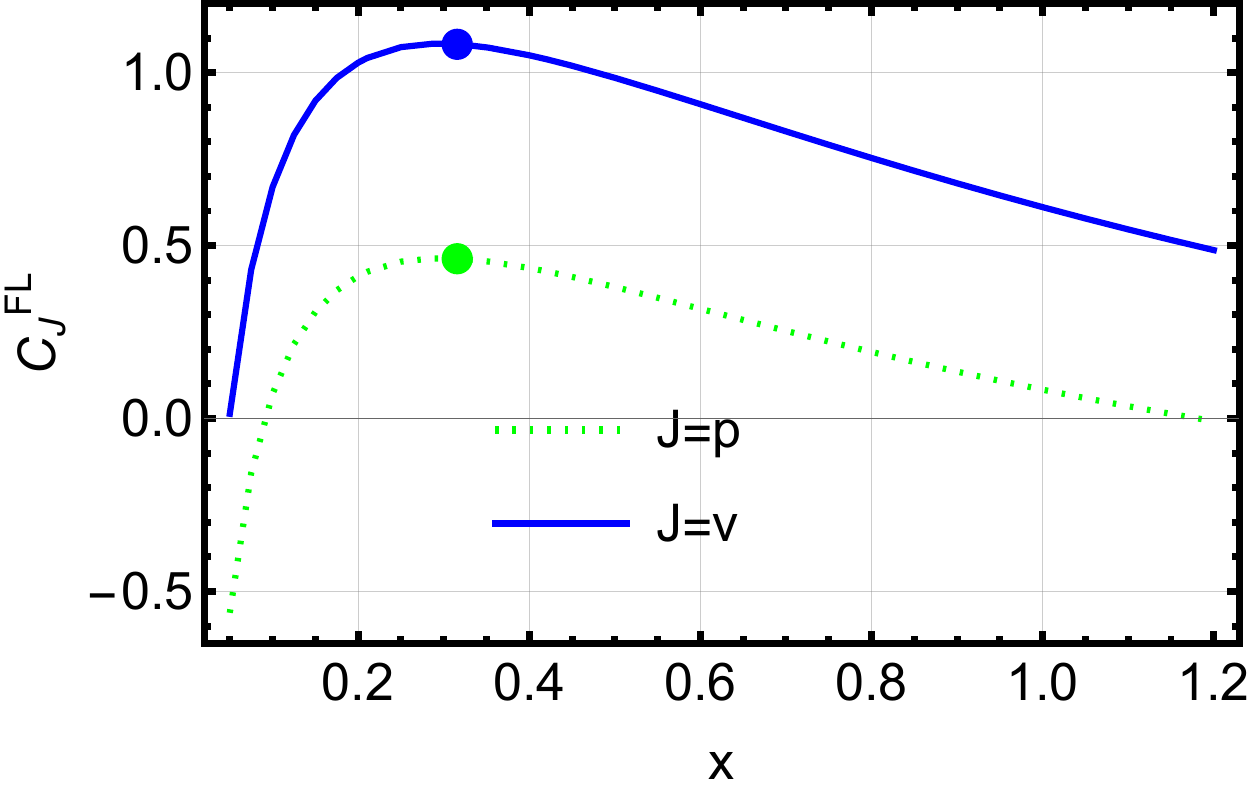}
	\caption{The same as Fig.~\ref{fig:CFFx}, but for  the profile of sub-coefficient $\mathcal{C}^{FL}_{J}(x)$ which has a pre-factor  $C_F  T_F n_l$. }
	\label{fig:CFLx}
\end{figure}

\begin{figure}[thb]
	\centering
	\includegraphics[width=0.45\textwidth]{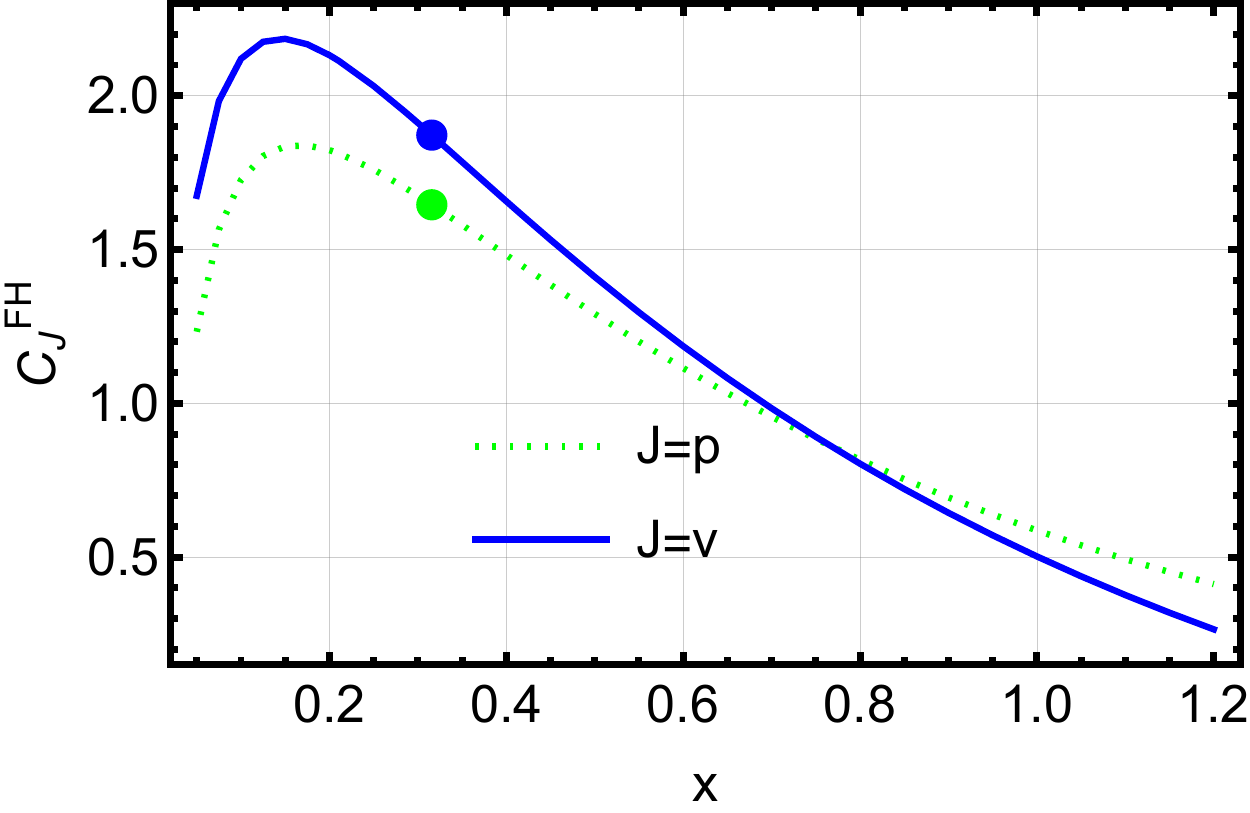}
	\caption{The same as Fig.~\ref{fig:CFFx}, but for  the profile of sub-coefficient $\mathcal{C}^{FH}_{J}(x)$ which has a pre-factor $C_F  T_F$. }
	\label{fig:CFHx}
\end{figure}

In the following, we will give the numerical results for the matching coefficients $\mathcal{C}_J(\mu_f,\mu,m_b,x)$ with $J=p,~v$ at NNLO accuracy.
The sub-coefficients $\mathcal{C}^{FF}_{J}$, $\mathcal{C}^{FA}_{J}$, $\mathcal{C}^{FL}_{J}$, and $\mathcal{C}^{FH}_{J}$ classified by different color/flavor structures in Eq.~\eqref{C0express}  are functions which only depend on the heavy quark mass ratio with $x=\frac{m_c}{m_b}$.  At physical heavy quark mass ratio, i.e., $x=x_0=\frac{1.5}{4.75}$~\cite{Qiao:2012hp,Qiao:2012vt,Qiao:2011zc}, we obtained the following highly accurate  numerical results  with about 30-digit precision for all four kinds of sub-coefficients.
\begin{align}
&\mathcal{C}^{FF}_{p}(x_0)=-18.1856109097151570253549607713,
\\&
\mathcal{C}^{FA}_{p}(x_0)=-11.9709902751165587736864132992,
\\&
\mathcal{C}^{FL}_{p}(x_0)=0.461971258745060837844427133019,
\\&
\mathcal{C}^{FH}_{p}(x_0)=1.64553283592627680478382129760.
\\&
\mathcal{C}^{FF}_{v}(x_0)=-15.8653228579431784031838005865,
\\&
\mathcal{C}^{FA}_{v}(x_0)=-11.0678680506800630685188604612,
\\&
\mathcal{C}^{FL}_{v}(x_0)=1.08196339731790945235792891668,
\\&
\mathcal{C}^{FH}_{v}(x_0)=1.87201601140852309779426933441.
\end{align}
In order to investigate the heavy quark mass dependence of the matching coefficients, we vary the heavy quark mass ratio $x$
 from $x_{min}=0.05$ to $x_{max}=1.2$. And we plotted the the heavy quark mass ratio $x$  dependence for the sub-coefficients $\mathcal{C}^{FF}_{J}$, $\mathcal{C}^{FA}_{J}$, $\mathcal{C}^{FL}_{J}$, and $\mathcal{C}^{FH}_{J}$   in Fig.~\ref{fig:CFFx}, Fig.~\ref{fig:CFAx}, Fig.~\ref{fig:CFLx}, and Fig.~\ref{fig:CFHx}, respectively. In these diagrams, $J=p$ represents the sub-coefficient for the pseudoscalar current while
$J=v$ represents the sub-coefficient for the vector current. From the curves in Figs.~(\ref{fig:CFFx}-\ref{fig:CFHx}), one can see the sub-coefficients are close to each other for both pseudoscalar and vector currents, except
$\mathcal{C}^{FL}_{J}$. The sub-coefficients $\mathcal{C}^{FF}_{J}$ and $\mathcal{C}^{FA}_{J}$ increase gradually  with the increase of the heavy quark mass ratio, while  $\mathcal{C}^{FL}_{J}$ and $\mathcal{C}^{FH}_{J}$ first increase and then reduce with the increase of the heavy quark mass ratio.

Fixing the renormalization scale $\mu=m_b=4.75\mathrm{GeV}$, $m_c=1.5\mathrm{GeV}$, and setting the factorization scale $\mu_f=1.2\mathrm{GeV}$, Eq.~\eqref{C0express} then reduces to
\begin{align}
&\mathcal{C}_p(\mu_f=1.2\mathrm{GeV},x=x_0)=
\nonumber\\&
1-1.40061\frac{\alpha_s^{(n_l=3)}(m_b)}{\pi}-30.69707\left(\frac{\alpha_s^{(n_l=3)}(m_b)}{\pi}\right)^2+\mathcal{O}(\alpha^3_s),
\\&
\mathcal{C}_v(\mu_f=1.2\mathrm{GeV},x=x_0)=
\nonumber\\&
1-2.06727\frac{\alpha_s^{(n_l=3)}(m_b)}{\pi}-33.56657\left(\frac{\alpha_s^{(n_l=3)}(m_b)}{\pi}\right)^2+\mathcal{O}(\alpha^3_s).
\end{align}

\begin{figure}[thb]
	\centering
	\includegraphics[width=0.45\textwidth]{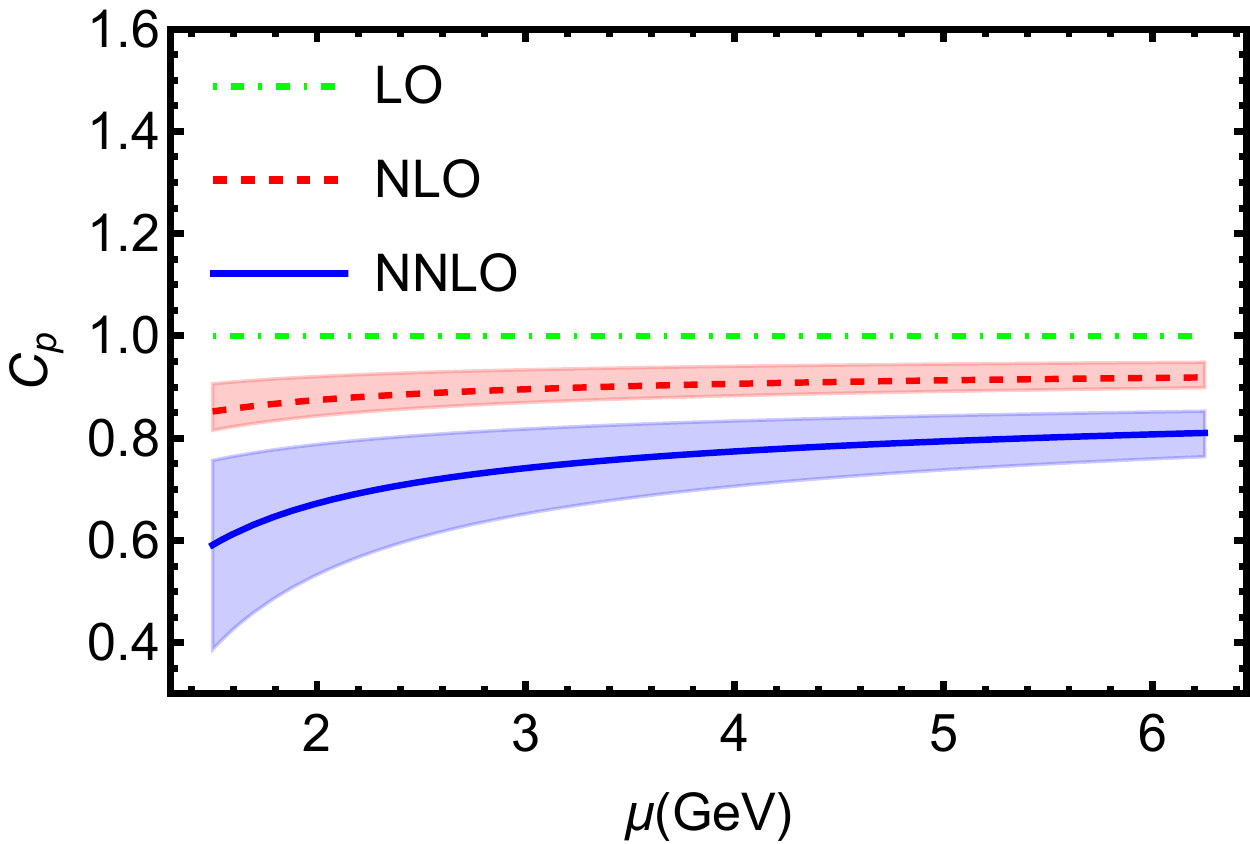}
	\caption{The renormalization scale dependence of the matching coefficient $\mathcal{C}_p$ for pseudoscalar current decay constant
 at LO,  NLO and  NNLO accuracy. The central values of  the matching coefficient $\mathcal{C}_p$ are calculated inputting the  physical values with $\mu_f=1.2\mathrm{GeV}$,  $m_b=4.75\mathrm{GeV}$ and $m_c=1.5\mathrm{GeV}$.   The error bands come from varying  $\mu_f$  from   1.5 to 1 $\mathrm{GeV}$, $m_b$  from   5.25 to 4.25 $\mathrm{GeV}$, $m_c$  from   2 to 1 $\mathrm{GeV}$. }
	\label{fig:Cpmudepand}
\end{figure}

\begin{figure}[thb]
	\centering
	\includegraphics[width=0.45\textwidth]{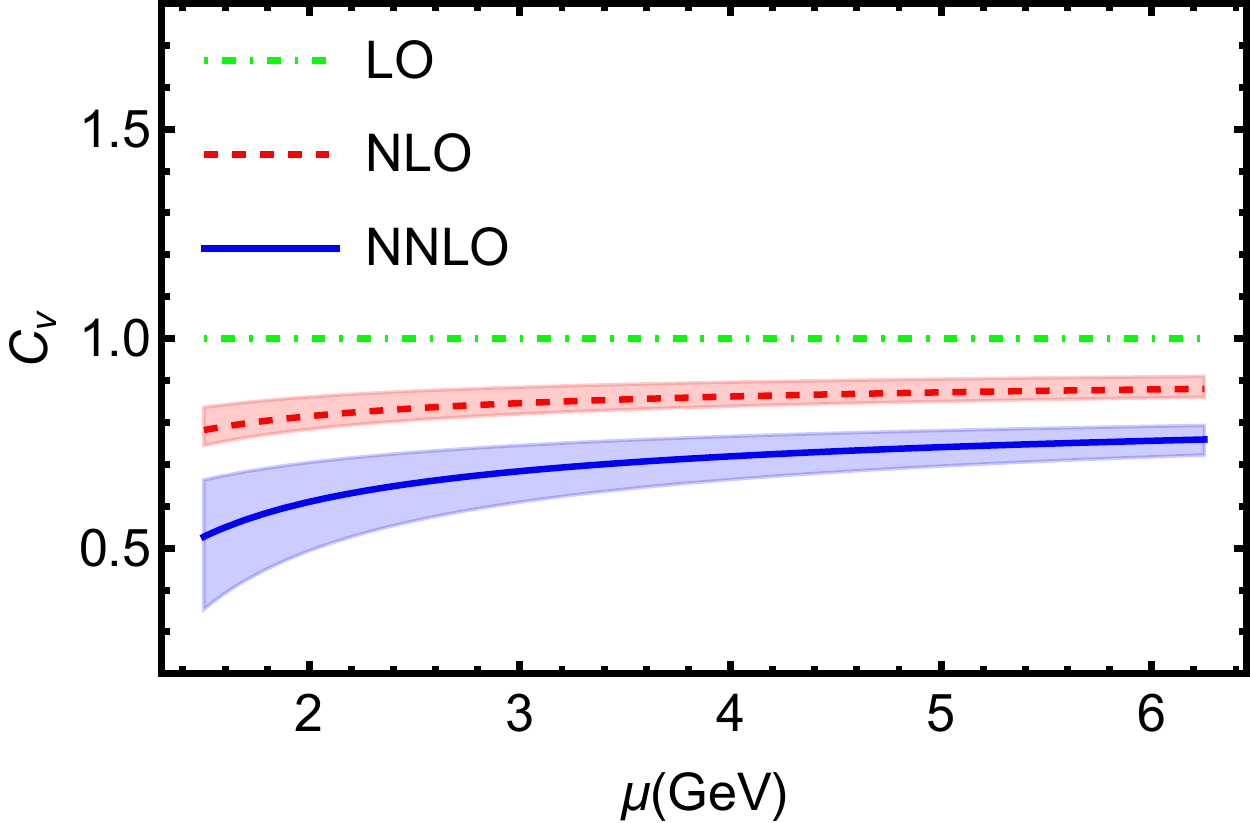}
	\caption{The same as Fig.~\ref{fig:Cpmudepand}, but for  the matching coefficient $\mathcal{C}_v$ for vector current decay constant. }
	\label{fig:Cvmudepand}
\end{figure}

After obtaining the value of $\alpha_s^{(n_l=3)}$ by Eq.~\eqref{aslexpress}, we present our numerical results of the matching coefficients for the pseudoscalar current and the vector current in Fig.~\ref{fig:Cpmudepand} and Fig.~\ref{fig:Cvmudepand}. From the figures, one can see the high-order corrections bring large scale-dependence in matching coefficients.
 This can be understood because the leading-order of matching coefficients is renormalized to 1.
 The scale-dependence in NLO corrections only rely  on strong coupling constant at $\alpha_s(\mu)$. The scale-dependence in NNLO corrections not only rely  on
 The strong coupling constant at $\alpha^2_s(\mu)$ but also the residue terms at order of $\alpha^2_s(\mu) \ln \mu^2$ which are not small for low scales.
 We summarize the results of the matching coefficients $\mathcal{C}_J$ for pseudoscalar and vector current decay constants
 at LO,  NLO and  NNLO accuracy in Tab.~\ref{tab:C0num}, where the uncertainties from all the parameters are included. From Tab.~\ref{tab:C0num},
 the two largest uncertainties are from the factorization scale $\mu_f$ and the renormalization factor $\mu$ at NNLO, while
 the uncertainties from the bottom and charm quark are relatively small.

\begin{table}[thb]
	\begin{center}
		\caption{The matching coefficients for both pseudoscalar and vector currents up to NNLO.
The central values of  the matching coefficient $\mathcal{C}_J$ are calculated inputting the  physical values with $\mu_f=1.2\mathrm{GeV}$, $\mu=4.75\mathrm{GeV}$, $m_b=4.75\mathrm{GeV}$ and $m_c=1.5\mathrm{GeV}$.
The errors are estimated by varying $\mu_f$  from   1.5 to 1 $\mathrm{GeV}$, $\mu$  from   6.25 to 3 $\mathrm{GeV}$, $m_b$  from   5.25 to 4.25 $\mathrm{GeV}$, and $m_c$  from   2 to 1 $\mathrm{GeV}$, respectively.}
		\label{tab:C0num}
\renewcommand\arraystretch{1.5}
		\begin{tabular}{ c c c c}
 \hline\hline
		             & LO         &  NLO                   & NNLO
  \\\hline
	$\mathcal{C}_p$	 & $1$ & $0.9117^{-0+0.0072+0.0061-0.0156}_{+0-0.0160-0.0064+0.0263}$   &    $0.7897^{-0.0310+0.0206+0.0119+0.0149}_{+0.0253-0.0482-0.0133-0.0141}$
     \\ \hline
	$\mathcal{C}_v$	 & $1$ &  $0.8697^{-0+0.0107+0.0061-0.0156}_{+0-0.0236-0.0064+0.0263}$  & $0.7363^{-0.0234+0.0230+0.0106+0.0117}_{+0.0191-0.0526-0.0117-0.0121}$
     \\		\hline \hline
		\end{tabular}
	\end{center}
\end{table}

Note that the explicit analytical expression for $\mathcal{C}_{a(\mu=0)}$ is given in Ref.~\cite{Chen:2015csa} at NNLO accuracy and the numerical results for $\mathcal{C}_{a(\mu=0)}$ is given
in Ref.~\cite{Feng:2022ruy} at NNNLO
accuracy. Our results are consistent with the previous results in Refs.~\cite{Onishchenko:2003ui,Chen:2015csa,Feng:2022ruy}.
Even though the QCD theory makes the two decay constants from pseudoscalar current and axial current(timelike-component) identical, i.e. $\mathcal{C}_p=\mathcal{C}_{a(\mu=0)}$.  Here we have explicit examined
this point by the independent calculation of the pseudoscalar decay constants.  On the other hand, the results for vector $B_c$ meson decay constant and its matching coefficient are novel.
 In the case of  $m_b=m_c=m_Q$, our result for the vector current agrees with the previous results in literatures~\cite{
Kniehl:2006qw,Egner:2022jot}.

For the pseudoscalar meson $B_c$ and vector meson $B_c^*$, the leptonic decay widths can be written as
\begin{align}
& \Gamma(B^+_c \to l^{+} + {\nu}_{l})=\frac{{|V_{bc}|}^2}{8\pi}  G_F^2 m_{B_c} m_l^2 \left(1-\frac{m_l^2}{m_{B_c}^2}\right)^2 {f^p_{B_c}}^2,
\\&
\Gamma(B_c^{*+} \to l^{+} + {\nu}_{l})=\frac{{|V_{bc}|}^2}{12\pi}  G_F^2 m_{B_c^*}^3 \left(1-\frac{m_l^2}{m_{B_c^*}^2}\right)^2 \left(1+\frac{m_l^2}{2 m_{B_c^*}^2}\right) {f^v_{B_c^*}}^2.
\end{align}
To evaluate the two decay constants  $f^p_{B_c}$ and $f^v_{B^*_c}$, we substitute the LDMEs in Eq.~\eqref{formula} with
\begin{align}
\left\langle 0\left|\chi^\dag_b \mathbf{\sigma}\psi_c \right| B^*_c(\mathbf{P})\right\rangle\approx
\left\langle 0\left|\chi^\dag_b \psi_c \right| B_c(\mathbf{P})\right\rangle\approx\sqrt{2 N_c} \; \psi_{B_c}(0),
\end{align}
where  $\psi_{B_c}(0)$ is the Schr\"{o}dinger   wave function at the origin for $B_c$ system and is predicted in potential models~\cite{Feng:2022ruy,Eichten:1995ch,Kiselev:2000jc,Ikhdair:2003ry,Shen:2021dat} as
\begin{align}
|\psi_{B_c}(0)|^2 \simeq [0.10,0.13] \mathrm{GeV}^3.
\end{align}
The values of input parameters are extracted from the latest PDG group~\cite{Workman:2022ynf} as following:
\begin{align}
&V_{bc}=0.0408,~ G_F=1.16638 \times 10^{-5} \mathrm{GeV}^{-2},
\nonumber\\&
m_e=0.000510999 \mathrm{GeV}, ~m_{\mu}=0.10566 \mathrm{GeV},
\nonumber\\&
m_{\tau}=1.777 \mathrm{GeV}, ~m_{B_c}=6.274 \mathrm{GeV}, ~\tau_{B_c}=0.51 \times 10^{-12} \mathrm{s}.\nonumber
\end{align}
For $B^*_c$, there are many theoretical predictions for its mass and decay widths~\cite{Godfrey:2004ya,Wang:2022cxy,Sun:2022hyk}. We use the following values in Refs.~\cite{Godfrey:2004ya,Zhou:2017svh}
\begin{align}
& m_{B_c^*}=6.314 \mathrm{GeV},~~
 \Gamma_{\mathrm{tot}}(B_c^*)=0.08 \times 10^{-6} \mathrm{GeV}.\nonumber
\end{align}

With the values of above input parameters, we  present our predictions to $B_c$ and $B_c^*$ decay constants  in Tab.~\ref{tab:fBcs}.
\begin{table}[thb]
	\begin{center}
		\caption{The predicted  decay constants  $f^p_{B_c}$ and $f^v_{B^*_c}$ up to NNLO.
The central values  are calculated inputting the  physical values with  $\psi_{B_c}(0)=\sqrt{0.12}\mathrm{GeV}^{\frac{3}{2}}$, $\mu_f=1.2\mathrm{GeV}$, $\mu=4.75\mathrm{GeV}$, $m_b=4.75\mathrm{GeV}$ and $m_c=1.5\mathrm{GeV}$.
The errors are estimated by varying $\psi_{B_c}(0)$  from $\sqrt{0.13}$ to $\sqrt{0.10}$ $\mathrm{GeV}^{\frac{3}{2}}$, $\mu_f$  from   1.5 to 1 $\mathrm{GeV}$, $\mu$  from   6.25 to 3 $\mathrm{GeV}$, $m_b$  from   5.25 to 4.25 $\mathrm{GeV}$, and $m_c$  from   2 to 1 $\mathrm{GeV}$, respectively.}
		\label{tab:fBcs}
\renewcommand\arraystretch{1.5}
		\begin{tabular}{ c c c} \hline\hline
              & $\frac{f^p_{B_c}}{10^{-1}\mathrm{GeV}}$   &  $\frac{f^v_{B^*_c}}{10^{-1}\mathrm{GeV}}$     \\\hline
		LO    & $4.79^{+0.20+0+0+0+0}_{-0.42-0-0-0-0}$                   &  $4.78^{+0.20+0+0+0+0}_{-0.42-0-0-0-0}$                                \\ \hline
		NLO   & $4.37^{+0.18+0+0.03+0.03-0.07}_{-0.38-0-0.08-0.03+0.13}$  &  $4.15^{+0.17+0+0.05+0.03-0.07}_{-0.36-0-0.11-0.03+0.13}$                               \\ \hline
		NNLO  & $3.78^{+0.15-0.15+0.10+0.06+0.07}_{-0.33+0.12-0.23-0.06-0.07}$ &$3.52^{+0.14-0.11+0.11+0.05+0.06}_{-0.31+0.09-0.25-0.06-0.06}$                        \\
			\hline \hline
		\end{tabular}
	\end{center}
\end{table}

Then we present our predictions to $B_c$ and $B_c^*$ leptonic decay widths in Tab.~\ref{tab:gammae}, Tab.~\ref{tab:gammamu}, and Tab.~\ref{tab:gammatau},  as well as the corresponding branching ratios in Tab.~\ref{tab:Brs}.

\begin{table}[thb]
	\begin{center}
		\caption{The predicted  leptonic decay width $\Gamma(B_c^+ \to e^{+} + {\nu}_{e})$  and $\Gamma(B_c^{+*} \to e^{+} + {\nu}_{e})$ up to NNLO.
The central values  are calculated inputting the  physical values with  $|\psi_{B_c}(0)|^2=0.12\mathrm{GeV}^3$, $\mu_f=1.2\mathrm{GeV}$, $\mu=4.75\mathrm{GeV}$, $m_b=4.75\mathrm{GeV}$ and $m_c=1.5\mathrm{GeV}$.
The errors are estimated by varying $|\psi_{B_c}(0)|^2$  from   0.13 to 0.10 $\mathrm{GeV}^3$, $\mu_f$  from   1.5 to 1 $\mathrm{GeV}$, $\mu$  from   6.25 to 3 $\mathrm{GeV}$, $m_b$  from   5.25 to 4.25 $\mathrm{GeV}$, and $m_c$  from   2 to 1 $\mathrm{GeV}$, respectively.}
		\label{tab:gammae}
\renewcommand\arraystretch{1.5}
		\begin{tabular}{ c c c} \hline\hline
              & $\frac{\Gamma(B^+_c \to e^{+} + {\nu}_{e})}{10^{-21}\mathrm{GeV}}$   &  $\frac{\Gamma(B_c^{*+} \to e^{+} + {\nu}_{e})}{10^{-13}\mathrm{GeV}}$     \\\hline
		LO    & $3.39^{+0.28+0+0+0+0}_{-0.56-0-0-0-0}$                   &  $3.45^{+0.29+0+0+0+0}_{-0.57-0-0-0-0}$                                \\ \hline
		NLO   & $2.82^{+0.23+0+0.04+0.04-0.10}_{-0.47-0-0.10-0.04+0.16}$  &  $2.61^{+0.22+0+0.06+0.04-0.09}_{-0.43-0-0.14-0.04+0.16}$                               \\ \hline
		NNLO  & $2.11^{+0.18-0.16+0.11+0.06+0.08}_{-0.35+0.14-0.25-0.07-0.07}$ &$1.87^{+0.16-0.12+0.12+0.05+0.06}_{-0.31+0.10-0.26-0.06-0.06}$                        \\
			\hline \hline
		\end{tabular}
	\end{center}
\end{table}

\begin{table}[thb]
	\begin{center}
		\caption{The same as Tab.~\ref{tab:gammae}, but for   $\Gamma(B_c^+ \to \mu^{+} + {\nu}_{\mu})$  and $\Gamma(B_c^{*+} \to \mu^{+} + {\nu}_{\mu})$.}
		\label{tab:gammamu}
\renewcommand\arraystretch{1.5}
		\begin{tabular}{ c c c} \hline\hline
              & $\frac{\Gamma(B_c \to \mu^{+} + {\nu}_{\mu})}{10^{-16}\mathrm{GeV}}$   &  $\frac{\Gamma(B_c^* \to \mu^{+} + {\nu}_{\mu})}{10^{-13}\mathrm{GeV}}$     \\\hline
		LO    & $1.45^{+0.12+0+0+0+0}_{-0.24-0-0-0-0}$                   &  $3.45^{+0.29+0+0+0+0}_{-0.57-0-0-0-0}$                                \\ \hline
		NLO   & $1.20^{+0.10+0+0.02+0.02-0.04}_{-0.20-0-0.04-0.02+0.07}$  &  $2.61^{+0.22+0+0.06+0.04-0.09}_{-0.43-0-0.14-0.04+0.16}$                               \\ \hline
		NNLO  & $0.90^{+0.08-0.07+0.05+0.03+0.03}_{-0.15+0.06-0.11-0.03-0.03}$ &$1.87^{+0.16-0.12+0.12+0.05+0.06}_{-0.31+0.10-0.26-0.06-0.06}$                        \\
			\hline \hline
		\end{tabular}
	\end{center}
\end{table}

\begin{table}[thb]
	\begin{center}
		\caption{The same as Tab.~\ref{tab:gammae}, but for   $\Gamma(B_c^+ \to \tau^{+} + {\nu}_{\tau})$  and $\Gamma(B_c^{*+} \to \tau^{+} + {\nu}_{\tau})$.}
		\label{tab:gammatau}
\renewcommand\arraystretch{1.5}
		\begin{tabular}{ c c c} \hline\hline
              & $\frac{\Gamma(B_c^+ \to \tau^{+} + {\nu}_{\tau})}{10^{-14}\mathrm{GeV}}$   &  $\frac{\Gamma(B_c^{*+} \to \tau^{+} + {\nu}_{\tau})}{10^{-13}\mathrm{GeV}}$     \\\hline
		LO    & $3.47^{+0.29+0+0+0+0}_{-0.58-0-0-0-0}$                   &  $3.04^{+0.25+0+0+0+0}_{-0.51-0-0-0-0}$                                \\ \hline
		NLO   & $2.88^{+0.24+0+0.05+0.04-0.10}_{-0.48-0-0.10-0.04+0.17}$  &  $2.30^{+0.19+0+0.06+0.03-0.08}_{-0.38-0-0.12-0.03+0.14}$                               \\ \hline
		NNLO  & $2.16^{+0.18-0.17+0.11+0.07+0.08}_{-0.36+0.14-0.26-0.07-0.08}$ &$1.65^{+0.14-0.10+0.10+0.05+0.05}_{-0.27+0.09-0.23-0.05-0.05}$                        \\
			\hline \hline
		\end{tabular}
	\end{center}
\end{table}

%\begin{table}[thb]
%	\begin{center}
%		\caption{The predictions of leptonic branching ratios of $B_c$ and $B_c^*$ at NNLO.
%The central values  are calculated inputting the  physical values with  $|\psi_{B_c}(0)|^2=0.12\mathrm{GeV}^3$, $\mu_f=1.2\mathrm{GeV}$, $\mu=4.75\mathrm{GeV}$, $m_b=4.75\mathrm{GeV}$ and $m_c=1.5\mathrm{GeV}$.
%The total errors come from varying $|\psi_{B_c}(0)|^2$  from  0.13 to 0.10  $\mathrm{GeV}^3$, $\mu_f$  from   1.5 to 1 $\mathrm{GeV}$, $\mu$  from   6.25 to 3 $\mathrm{GeV}$, $m_b$  from   5.25 to 4.25 $\mathrm{GeV}$, and $m_c$  from   2 to 1 $\mathrm{GeV}$.}
%		\label{tab:Brs}
%\renewcommand\arraystretch{1.5}
%\tabcolsep=0.2cm
%		\begin{tabular}{ c c c }
% \hline\hline
%		                     & $B_c$                             &  $B_c^*$
%  \\\hline
%	$e^{+}{\nu}_{e}$	     & $\left(1.64^{+0.44}_{-0.71}\right)\times 10^{-9}$   &  $\left(2.34^{+0.61}_{-1.01}\right)\times 10^{-6}$
%     \\ \hline
%	$\mu^{+}{\nu}_{\mu}$	 & $\left(7.00^{+1.89}_{-3.01}\right)\times 10^{-5}$   &  $\left(2.34^{+0.61}_{-1.01}\right)\times 10^{-6}$
%     \\ \hline
%	$\tau^{+}{\nu}_{\tau}$	 & $\left(1.68^{+0.45}_{-0.72}\right)\times 10^{-2}$   &  $\left(2.06^{+0.54}_{-0.89}\right)\times 10^{-6}$
%     \\		\hline \hline
%		\end{tabular}
%	\end{center}
%\end{table}

\begin{table}[thb]
	\begin{center}
		\caption{The predictions of leptonic branching ratios of $B_c$ and $B_c^*$ at NNLO.
The central values  are calculated inputting the  physical values with  $|\psi_{B_c}(0)|^2=0.12\mathrm{GeV}^3$, $\mu_f=1.2\mathrm{GeV}$, $\mu=4.75\mathrm{GeV}$, $m_b=4.75\mathrm{GeV}$ and $m_c=1.5\mathrm{GeV}$.
The total errors come from varying $|\psi_{B_c}(0)|^2$  from   0.13 to 0.10 $\mathrm{GeV}^3$, $\mu_f$  from   1.5 to 1 $\mathrm{GeV}$, $\mu$  from   6.25 to 3 $\mathrm{GeV}$, $m_b$  from   5.25 to 4.25 $\mathrm{GeV}$, and $m_c$  from   2 to 1 $\mathrm{GeV}$.}
		\label{tab:Brs}
\renewcommand\arraystretch{1.5}
\tabcolsep=0.15cm
		\begin{tabular}{ c c c c}
 \hline\hline
		             & $e^{+}{\nu}_{e}$         &  $\mu^{+}{\nu}_{\mu}$                   & $\tau^{+}{\nu}_{\tau}$
  \\\hline
	$B_c$	 & $\left(1.64^{+0.44}_{-0.71}\right)\times 10^{-9}$ & $\left(7.00^{+1.89}_{-3.01}\right)\times 10^{-5}$   &    $\left(1.68^{+0.45}_{-0.72}\right)\times 10^{-2}$
     \\ \hline
	$B_c^*$	 & $\left(2.34^{+0.61}_{-1.01}\right)\times 10^{-6}$ &  $\left(2.34^{+0.61}_{-1.01}\right)\times 10^{-6}$  & $\left(2.06^{+0.54}_{-0.89}\right)\times 10^{-6}$
     \\		\hline \hline
		\end{tabular}
	\end{center}
\end{table}

From Tabs.~\ref{tab:gammae}, \ref{tab:gammamu}, and \ref{tab:gammatau}, the leptonic decay widths for $B^*_c$ are around $10^{-13}\mathrm{GeV}$,
while the electronic, muonic and tauonic  decay widths for $B_c$ are around $10^{-21}\mathrm{GeV}$, $10^{-16}\mathrm{GeV}$, $10^{-14}\mathrm{GeV}$, respectively.
From Tab.~\ref{tab:Brs},  one can see the leptonic branching ratios for $B^*_c$ are around $10^{-6}$.
The branching ratio for the tauonic decay of $B_c$ is around $10^{-2}$ while the branching ratio for the muonic decay of $B_c$ is around $10^{-4}-10^{-5}$ .

Consider the hadronic production of $B_c$  and $B^*_c$ has a large uncertainty and their cross-sections at LHC are from tens to hundreds nanobarn~\cite{Chang:2003cq,Chang:2003cr,Chang:2005hq},
there are tens to hundreds  $B_c^{*+} \to l^{+} + {\nu}_{l}$ events, while hundreds to thousands $B_c^{+} \to \mu^{+} + {\nu}_{\mu}$ events at LHC for $1fb^{-1}$ proton proton collision data at 14TeV. Of course, the branching ratio of $B_c^{+} \to \tau^{+} + {\nu}_{\tau}$ is around 3 order of the branching ratio of $B_c^{+} \to \mu^{+} + {\nu}_{\mu}$, thus this channel
shall be also a good detect channel of $B_c$ meson if the reconstruction of tauon lepton is well-controlled.  In total,
we expect these leptonic decay channels for both $B_c$  and $B^*_c$ can be accessible at LHC precision experiments.

\section{Conclusion}
In this paper, we have performed a NNLO calculation of the decay constants of beauty-charmed meson $B_{c}$ and $B^*_{c}$.
The NNLO result for vector current decay constant is novel. The updated leptonic decay branching ratios combined with
the latest extraction of NRQCD LDMEs of $B_c$ meson shall be tested in future experiments. Through the
careful studies of the decay constants of $B_c$ meson, one can expect that more and more decay channels of beauty-charmed mesons
are accessible and their absolute branching ratios can be measured. The novel results of the anomalous dimension for the
vector current in NRQCD shall provide more information on the renormalization properties of the NRQCD LDMEs. The NNLO matching
coefficients are also helpful to investigate the behaviours when the doubly heavy quarks are in their threshold region.

%%%%%%%%%%%%%%%%%%%%%%%%%%%%%%%%%%%%%%%%%%%%%%%%%%%%%%%%%%%%%%%%%%%%%%%%%%%%%%%%
\section*{Acknowledgements}
%%%%%%%%%%%%%%%%%%%%%%%%%%%%%%%%%%%%%%%%%%%%%%%%%%%%%%%%%%%%%%%%%%%%%%%%%%%%%%%%
We thank L.~B. Chen, Y.~M. Li, X. Liu, W.~L. Sang and C.~Y. Wang for many useful discussions. This work is supported by NSFC under grant No.~11775117 and No.~12075124,  and by Natural Science Foundation of Jiangsu under Grant No.~BK20211267.

%%%%%%%%%%%%%%%%%%%%%%%%%%%%%%%%%%%%%%%%%%%%%%

%%%%%%%%%%%%%%%%%%%%%%%%%%%%%%%%%%%%%%%%%%%%%%%%%%%%%%%%%%%%%%%%%%%%%%%%%%%%%%%%
\section*{Appendix}
%%%%%%%%%%%%%%%%%%%%%%%%%%%%%%%%%%%%%%%%%%%%%%%%%%%%%%%%%%%%%%%%%%%%%%%%%%%%%%%%

\begin{appendix}

\begin{widetext}

Allowing $n_b$ quarks with mass $m_b$, $n_c$ quarks  with mass $m_c$ and $n_l$ massless quarks appearing in the quark loop, bottom quark on-shell wave function renormalization constant up to NNLO reads
\begin{align}
Z_{2,b}
&=
1
+
\frac{ \alpha _s}{\pi } C_F\bigg\{-\frac{3 }{4   \epsilon }
-\frac{3}{4} \ln \frac{\mu ^2}{m_b^2}-	1
	-\frac{ \epsilon }{16  }\left(6 \ln^2 \frac{\mu ^2}{m_b^2}+16
	\ln \frac{\mu ^2}{m_b^2}+\pi ^2+32\right)
\nonumber\\	&
-  \epsilon ^2\bigg[\frac{1}{8} \ln^3 \frac{\mu ^2}{m_b^2}+\frac{1}{2} \ln^2 \frac{\mu
		^2}{m_b^2}+\left(2+\frac{\pi ^2}{16}\right) \ln \frac{\mu ^2}{m_b^2}-\frac{\zeta _3}{4}+\frac{\pi ^2}{12}+4\bigg]\bigg\}
\nonumber\\ &
+
 \frac{\alpha _s^2}{ \pi ^2}C_F\bigg\{
 C_F\bigg[\frac{9}{32 \epsilon ^2}+\frac{1}{192 \epsilon }\left(108  \ln \frac{\mu ^2}{m_b^2}+153 \right)+
\frac{9}{16} \ln^2 \frac{\mu
		^2}{m_b^2}+\frac{51}{32} \ln \frac{\mu ^2}{m_b^2}+\pi ^2 \ln 2-\frac{3
		\zeta _3}{2}-\frac{49 \pi ^2}{64}+\frac{433}{128}\bigg]
	\nonumber\\		&
+C_A \bigg[\frac{11}{32 \epsilon ^2}-\frac{127}{192 \epsilon} -\frac{11}{32} \ln^2 \frac{\mu ^2}{m_b^2}-\frac{215}{96} \ln \frac{\mu ^2}{m_b^2}-\frac{1}{2} \pi ^2 \ln 2+\frac{3 \zeta
		_3}{4}+\frac{5 \pi ^2}{16}-\frac{1705}{384}\bigg]
	\nonumber\\	&
+ T_F  n_b\bigg[\frac{1}{16 \epsilon }\left(4 \ln \frac{\mu ^2}{m_b^2} +1 \right)+\frac{3}{8} \ln^2 \frac{\mu
		^2}{m_b^2}+\frac{11}{24} \ln \frac{\mu ^2}{m_b^2}-\frac{5 \pi ^2}{16}+\frac{947}{288}\bigg]
		\nonumber\\		&
+T_F  n_c \bigg[\frac{1}{16 \epsilon }\left(4 \ln
		\frac{\mu ^2}{m_b^2}-8  \ln x+1\right)+\frac{\pi ^2 x^4}{4}-\frac{5 \pi ^2 x^3}{8 }+\frac{7 x^2}{4}-\frac{3
		\pi ^2 x}{8}	+\frac{3}{8} \ln^2 \frac{\mu ^2}{m_b^2}+\frac{11}{24} \ln \frac{\mu ^2}{m_b^2}
	\nonumber\\		&
	+\ln (x) \ln \left(x+1\right)\left(-\frac{3 x^4}{2}-\frac{5 x^3}{4}-\frac{3 x}{4 }-\frac{1}{2}\right) +\text{Li}_2\left(x\right) \left(-\frac{3 x^4}{2}+\frac{5 x^3}{4 }+\frac{3 x}{4}-\frac{1}{2}\right)
	\nonumber\\		&
	+\ln^2\left( x\right) \left(\frac{3 x^4}{2 }+1\right)- \text{Li}_2\left(-x\right)\left(\frac{3 x^4}{2}+\frac{5 x^3}{4}+\frac{3 x}{4}+\frac{1}{2}\right)
			\nonumber\\		&
	+\ln\left(x\right)\left(x^2-\ln \frac{\mu ^2}{m_b^2}-\ln \left(1-x\right) \left(\frac{3 x^4}{2}-\frac{5 x^3}{4}-\frac{3 x}{4}+\frac{1}{2}\right)+\frac{2}{3}\right) +\frac{5 \pi^2}{48}+\frac{443}{288}\bigg]
			\nonumber\\		&
	+
T_F  n_l \bigg[\frac{ -1 }{8 \epsilon ^2}+	\frac{11}{48 \epsilon }
	+	
	\frac{1}{8} \ln^2 \frac{\mu ^2}{m_b^2} +\frac{19}{24} \ln  \frac{\mu ^2}{m_b^2} +\frac{\pi
		^2}{12}+\frac{113}{96}\bigg]
	\bigg\}.
\end{align}

And allowing $n_b$ quarks with mass $m_b$, $n_c$ quarks  with mass $m_c$ and $n_l$ massless quarks appearing in the quark loop, charm quark on-shell wave function renormalization constant up to NNLO    can be obtained as
\begin{align}
&Z_{2,c}={Z_{2,b}}|_{\, m_b\rightarrow m_c;\, x\rightarrow \frac{1}{x};\,  n_b\leftrightarrow n_c}.
\end{align}

 Allowing $n_b$ quarks with mass $m_b$, $n_c$ quarks  with mass $m_c$ and $n_l$ massless quarks appearing in the quark loop, bottom quark on-shell mass renormalization constant up to NNLO reads
\begin{align}
Z_{m,b}
&=
1
+
\frac{ \alpha _s}{\pi } C_F\bigg\{-\frac{3 }{4   \epsilon }
-\frac{3}{4} \ln \frac{\mu ^2}{m_b^2}-	1
-\frac{ \epsilon }{16  }\left(6 \ln^2 \frac{\mu ^2}{m_b^2}+16
\ln \frac{\mu ^2}{m_b^2}+\pi ^2+32\right)
\nonumber\\	&
-  \epsilon ^2\bigg[\frac{1}{8} \ln^3 \frac{\mu ^2}{m_b^2}+\frac{1}{2} \ln^2 \frac{\mu
	^2}{m_b^2}+\left(2+\frac{\pi ^2}{16}\right) \ln \frac{\mu ^2}{m_b^2}-\frac{\zeta _3}{4}+\frac{\pi ^2}{12}+4\bigg]\bigg\}
\nonumber\\ &
+\frac{\alpha _s^2}{\pi ^2} C_F\bigg\{
 C_F\bigg[
\frac{9}{32  \epsilon ^2}+
\frac{1}{192	\epsilon } \left(108  \ln \frac{\mu ^2}{m_b^2}+135 \right)+
 \frac{9}{16} \ln^2 \frac{\mu ^2}{m_b^2}+\frac{45}{32} \ln \frac{\mu ^2}{m_b^2}+\frac{1}{2} \pi ^2 \ln 2-\frac{3 \zeta_3}{4}-\frac{17 \pi
		^2}{64}+\frac{199}{128}\bigg]
		\nonumber\\		&
			+
C_A \bigg[
\frac{11}{32  \epsilon ^2}-
		\frac{97}{192	\epsilon }
		-\frac{11}{32} \ln^2 \frac{\mu ^2}{m_b^2}-\frac{185}{96} \ln \frac{\mu ^2}{m_b^2}-\frac{1}{4} \pi ^2 \ln 2+\frac{3 \zeta
		_3}{8}+\frac{\pi ^2}{12}-\frac{1111}{384}\bigg]
			\nonumber\\		&
	+
 T_F n_b	\bigg[
 \frac{-1}{8  \epsilon ^2}+
 \frac{5}{48	\epsilon }+
 \frac{1}{8} \ln^2 \frac{\mu ^2}{m_b^2}+\frac{13}{24} \ln \frac{\mu ^2}{m_b^2}-\frac{\pi ^2}{6}+\frac{143}{96}\bigg]
  	\nonumber\\		&
  + T_F   n_c \bigg[
  \frac{ -1}{8  \epsilon ^2}+
  \frac{5}{48	\epsilon } +
  \frac{ x^4}{2 }\ln^2  x+\frac{\pi ^2 x^4}{12}-\frac{\pi ^2 x^3}{4}+\frac{3 x^2}{4 }-\frac{\pi ^2 x}{4}+\frac{1}{8} \ln^2 \frac{\mu
		^2}{m_b^2}+\frac{13}{24} \ln \frac{\mu ^2}{m_b^2}
	  	\nonumber\\		&
	+\ln \left(x\right) \ln \left(x+1\right) \left(-\frac{x^4}{2}-\frac{x^3}{2 }-\frac{x}{2
	}-\frac{1}{2}\right)+\text{Li}_2\left(x\right) \left(-\frac{x^4}{2 }+\frac{x^3}{2 }+\frac{x}{2}-\frac{1}{2}\right)
		\nonumber\\		&
	-\text{Li}_2\left(-x\right) \left(\frac{x^4}{2
	}+\frac{x^3}{2}+\frac{x}{2}+\frac{1}{2}\right)+\ln \left(x\right) \left(\frac{x^2}{2 }-\ln \left(1-x\right) \left(\frac{x^4}{2}-\frac{x^3}{2
	}-\frac{x}{2}+\frac{1}{2}\right)\right)+\frac{\pi ^2}{12}+\frac{71}{96}\bigg]
		\nonumber\\		&
	+
 T_F    n_l	\bigg[
 \frac{-1 }{8  \epsilon ^2}
 + \frac{5}{48	\epsilon } +
  \frac{1}{8} \ln^2 \frac{\mu ^2}{m_b^2}+\frac{13}{24} \ln \frac{\mu ^2}{m_b^2}+\frac{\pi
		^2}{12}+\frac{71}{96}\bigg]
		\bigg\}.
\end{align}

And allowing $n_b$ quarks with mass $m_b$, $n_c$ quarks  with mass $m_c$ and $n_l$ massless quarks appearing in the quark loop, charm quark on-shell mass renormalization constant up to NNLO    can be obtained as
\begin{align}
&Z_{m,c}={Z_{m,b}}|_{\, m_b\rightarrow m_c;\, x\rightarrow \frac{1}{x};\,  n_b\leftrightarrow n_c}
\end{align}

\end{widetext}
\end{appendix}

\end{document}